\documentstyle[preprint,prl,aps]{revtex}
\newcommand{\sr}{\scriptstyle}
\newcommand{\ov}{\overline}
\begin{document}

\tightenlines

\draft

\title{Director configuration of planar solitons in nematic
liquid crystals}

\author{Henryk Arod\'z  and Joachim Stelzer}
\address{Instytut Fizyki, Uniwersytet Jagiello\'nski,
Reymonta 4, 30-059 Krak\'ow, Poland}
\maketitle
\widetext

\begin{abstract}
The director configuration of disclination lines in nematic liquid crystals
in the presence of an external magnetic field is evaluated.
Our method is a combination of
a polynomial expansion for the director and of further analytical 
approximations which are tested against a numerical shooting method.
The results are particularly simple when the elastic constants are equal, 
but we discuss the general  case of elastic anisotropy. The
director field is continuous everywhere apart from a straight
line segment whose length depends on the value of the magnetic field.
This indicates the possibility of an elongated defect core for disclination 
lines in nematics due to an external magnetic field.
\end{abstract}
\pacs{PACS:}
\narrowtext

\section{INTRODUCTION}

Nematic liquid crystals are systems which are positionally disordered,
but reveal a long-range orientational order 
\cite{Degennes1}. This property is described on a mesoscopic level by a
unit vector field ${\bbox{n}}({\bbox{r}})$, which
is called {\em director}. Due to the absence of permanent dipolar moments
in nematics the director just indicates the orientation, 
but its has neither head nor tail. This particular feature yields
very interesting defect structures in nematic liquid crystals. For instance,
the director field shows line defects in three dimensions (or,
equivalently,  point defects in two dimensions), called
{\em disclinations}, which have been studied and classified by topological
methods \cite{Toulouse,Mermin,Mineyev1,Trebin}. 
Unlike in spin systems, disclinations of topological charge
$\pm \frac{1}{2}$ are possible and stable in nematics. When an external
magnetic field is applied perpendicular to such a disclination line, 
the resulting director configuration becomes even more interesting --
it can be regarded as a domain wall filling a half-plane which terminates
in the disclination line. Such walls with edges, known as planar solitons 
\cite{Chandrasekhar}, have been discussed for superfluid Helium-3 by 
Mineyev and Volovik \cite{Mineyev2}. Whereas the qualitative behaviour 
of these soliton-like objects is well-established,
a quantitative understanding of their structure can be obtained only from
a thorough analysis of the underlying field theory. Its is
the aim of our paper to perform such calculations. Our approach is
based on a polynomial
expansion of the director field. This method has been used previously
both in relativistic field theories \cite{Arodz1,Arodz2}
and for the evaluation of domain wall dynamics in nematics \cite{Stelzer}.
It yields approximate analytical solutions for the director orientation.

The paper is organized as follows. In Section II the director field
equation for the planar solitons is derived.
Section III develops a method for obtaining an approximate solution
for the tilt angle of the director. Our technique is a
combination of the polynomial expansion \cite{Arodz1,Arodz2,Stelzer}
with further approximations which are tested by means of a 
numerical shooting method \cite{Numrec}. The discussion is performed within
the framework 
of the Oseen-Z\"ocher-Frank  elasticity  \cite{Oseen,Zocher,Frank}.  
In Section  IV we estimate the energy of the defect 
core of the planar solitons, 
and we minimize the total energy of the solitons in order to determine the
length of the core. Section V contains  concluding remarks.

\section{DIRECTOR FIELD EQUATION AND BOUNDARY CONDITIONS FOR PLANAR
SOLITONS}

\subsection{Free energy and field equation}

The geometry for planar solitons in nematic liquid crystals is
drawn schematically in Figs.~1 (positive soliton) and 2
(negative soliton). The director field is essentially planar,
perpendicular to a
disclination line of strength $\pm \frac{1}{2}$ along  the $z$ 
direction of a Cartesian coordinate frame. Because the structure is 
independent on $z$,  we restrict ourselves to the $x$-$y$ plane ($z = 0$). 
Now we impose a magnetic field in the plane of the director  along the $x$
axis. Due to the magnetic anisotropy of the nematic the director
tends to align along
the magnetic field. However, the topological charge of the disclination
has to be conserved. The resulting structure is a planar domain wall
of N\'eel type \cite{Neel,Helfrich,Stelzer}
which ends in the disclination line 
\cite{Mineyev2,Chandrasekhar}. Locally, close to
the disclination, the director field preserves the defect structure.
However, in a plane at a finite distance from the disclination line, 
which is given
by the half width $y_{0}$ of the planar N\'eel wall (Figs.~1 and 2), 
the director field is aligned parallel to the external magnetic field
\cite{Mineyev2,Chandrasekhar}.

Due to the translational symmetry along the $z$ axis it is sufficient
to perform the calculations
in two dimensions only. The director orientation is then completely
determined
by the {\em tilt angle field} $\Phi(x,y)$, which is measured with respect
to the direction of the magnetic field ${\bbox{H}}$ ($x$ axis),
\begin{equation} \label{director}
{\bbox{n}} = \cos\Phi(x,y)\,\hat{\bbox{x}}
+ \sin\Phi(x,y)\,\hat{\bbox{y}}, \qquad
{\bbox{H}} = H_0\,\hat{\bbox{x}}.
\end{equation}
We look for static  director configurations, hence $\Phi$ does not
depend on time. 

The static director orientation inside the soliton
corresponds to a configuration minimizing the total free energy $F$ (per unit
length in the $z$ direction) 
which contains both the energy of the nematic phase
$F_{\mathrm{nem}}$ and the core energy of the disclination 
$F_{\mathrm{core}}$. (Within the defect core local phase 
transitions may occur.) The nematic energy $F_{\mathrm{nem}}$ is the area
integral of a free energy density ${\cal F}_{\mathrm{nem}}$.
This free energy density, in turn, consists of  elastic contributions 
(due to distortions of the director
field) and of a magnetic part (taking into account the interaction of the
nematic with the external magnetic field), hence 
${\cal F}_{\mathrm{nem}} = {\cal F}_{\mathrm{elast}} 
+ {\cal F}_{\mathrm{mag}}$.
The elastic free energy density follows from the Oseen-Z\"ocher-Frank
expression \cite{Oseen,Zocher,Frank}.
\begin{equation} \label{Felast}
{\cal F}_{\mathrm{elast}} =
\frac{\sr 1}{\sr 2}\,K_{11}\,(\mbox{div}\,{\bbox{n}})^2
+ \frac{\sr 1}{\sr 2}\,K_{33}\,({\bbox{n}}\times
\mbox{curl}\,{\bbox{n}})^2.
\end{equation}
In (\ref{Felast}) $K_{11}$ and $K_{33}$ denote the elastic constants
for {\em splay} and {\em bend} deformations in the nematic.
Due to the restriction to planar director fields according to 
(\ref{director}) there are no {\em twist} deformations
and the elastic constant $K_{22}$ does not enter the
calculations.

The magnetic free energy density couples the director ${\bbox{n}}$ to
the magnetic field ${\bbox{H}}$ via the anisotropy of the magnetic
susceptibility $\Delta\chi$ ($\mu_0$ means the magnetic field constant).
\begin{equation} \label{Fmag}
{\cal F}_{\mathrm{mag}} = - \frac{\sr 1}{\sr 2}\,\mu_0\,\Delta\chi\,
({\bbox{n}}\cdot{\bbox{H}})^2.
\end{equation}
When inserting the ansatz for the planar director field 
(\ref{director}) into (\ref{Felast}) and
(\ref{Fmag}), we obtain the free energy density  
${\cal F}_{\mathrm{nem}}$ of the nematic phase.
\begin{eqnarray} \label{fed}
{\cal F}_{\mathrm{nem}} &=& \frac{\sr 1}{\sr 4}\,(K_{11} + K_{33})\,
(\Phi_x^2 + \Phi_y^2)
+ \frac{\sr 1}{\sr 4}\,(K_{33} - K_{11})\,
(\Phi_x^2 - \Phi_y^2)\,\cos 2\Phi 
\nonumber \\
& & + \frac{\sr 1}{\sr 2}\,(K_{33} - K_{11})\,
\Phi_x\,\Phi_y\,\sin 2\Phi
- \frac{\sr 1}{\sr 4}\,\mu_0\,\Delta\chi\,H_0^2\,(1 + \cos 2\Phi).
\end{eqnarray}
In (\ref{fed}) $\Phi_x$ and $\Phi_y$ denote partial derivatives of
the tilt angle with respect to the spatial coordinates.
The energy of the defect core $F_{\mathrm{core}}$ 
will be discussed separately in Section IV.

The director configuration for the planar soliton, which minimizes the 
energy of the nematic phase, follows as a solution of the corresponding
Euler-Lagrange equation
\begin{equation}
\frac{\delta {\cal F}_{\mathrm{nem}}}{\delta n_i}
\equiv 
\frac{\partial {\cal F}_{\mathrm{nem}}}{\partial n_i}
- \partial_j\,\left(\frac{\partial{\cal F}_{\mathrm{nem}}}{\partial\,
(\partial_j n_i)}\right) 
=0.
\end{equation}
The resulting equation for the tilt angle field $\Phi(x,y)$ can be written
in the following form
\begin{eqnarray} 
& & \Phi_{xx} + \Phi_{yy}
+ \overline{K}\, [ \partial_x\,(\Phi_x\,\cos 2\Phi)
- \partial_y\,(\Phi_y\,\cos 2\Phi) ]
\nonumber \\
& & + \overline{K}\, [ \partial_x\,(\Phi_y\,\sin 2\Phi)
+ \partial_y\,(\Phi_x\,\sin 2\Phi) ]
+ \overline{K}\, (\Phi_x^2 - \Phi_y^2 - 2 \Phi_x\Phi_y)\,\sin 2\Phi \nonumber \\
& & - \frac{\mu_0\,\Delta\chi\, H_0^2}{K_{11} +K_{33}} \,\sin 2\Phi
\qquad = 0,  \label{equ}
\end{eqnarray}
where
\begin{displaymath}
\overline{K} = \frac{K_{33} - K_{11}}{K_{11} + K_{33}}   
\end{displaymath}
is the elastic anisotropy.

\subsection{Boundary conditions}

The boundary conditions are an essential feature of the planar solitons.
As discussed above, the defect structure is surrounded by a homogeneous
director field and by a planar N\'eel wall. According to the
choice of our Cartesian coordinate frame (Figs.~1 and 2)
the tilt angle should be zero at $y=\pm y_{0}$,
where $y_{0}$ is the half width of the N\'eel wall. Additionally, it should
glue smoothly to the homogeneous orientation. Thus the
boundary conditions in the $y$ direction 
(perpendicular to the magnetic field) are given by
\begin{equation} \label{boundy}
\Phi(x,\,y=y_{0}) = 0, \qquad \Phi(x,\,y=-y_{0}) = \pm \pi, \qquad
\Phi_y(x,\,y=\pm y_{0}) = 0,
\end{equation}
where $\pm \pi$ is for the positive and negative soliton, respectively.

In the $x$ direction (parallel to the magnetic field),
the director field at $x\leq 0$ coincides with the planar
N\'eel wall. For increasing  $x$ coordinate the
domain wall structure is destroyed and the director
field changes towards the homogeneous orientation, parallel to the magnetic 
field, which is reached at $x_0$. Hence,
\begin{equation} 
\label{boundx}
\Phi(x=0,\,y) = \Phi_{\mathrm{Neel}}(y), \qquad
\Phi(x=x_0,\,y) = 0. 
\end{equation}
It is important to note that
the value of $x_0$ is yet unknown at this stage.

The function $\Phi_{\mathrm{Neel}}(y)$ describes the
director inversion  due to the planar N\'eel wall.
It can be determined by solving the field equation
(\ref{equ}) in one dimension. For $x\le 0$ there
is no dependence on the coordinate $x$ and the
field equation is simplified to
\begin{equation}  \label{equ1}
(1 - \overline{K}\,\cos 2\Phi) \Phi_{yy} 
+ \overline{K}\,\Phi_{y}^2\,\sin 2\Phi
- \frac{\mu_0\,\Delta\chi\,H_0^2}{K_{11}+K_{33}}\,\sin 2\Phi\, = 0.
\end{equation}

The center line of the cross section of the domain wall with the $x$-$y$
plane coincides with the line $x\leq 0,\,y=0$, with tilt
angle $\Phi= \pm \frac{\pi}{2}$ on it. Now, following the approach
developed in our recent publication \cite{Stelzer}, we apply a
{\em polynomial expansion} of the tilt angle
up to third order in the distance $y$ 
from the center line, 
\begin{equation} \label{poly1}
\Phi_{\mathrm{Neel}}(y) = \pm\frac{\pi}{2} 
\mp \frac{3\pi}{4}\,\frac{y}{y_0} 
\pm \frac{\pi}{4}\,\left(\frac{y}{y_0}\right)^3.
\end{equation}
The different signs are valid for positive and negative solitons,
respectively. 
Due to the choice of the coefficients in the expansion
(\ref{poly1}), the boundary conditions (\ref{boundy}) 
are fulfilled. With the approximate expression (\ref{poly1}) for 
$\Phi_{\mathrm{Neel}}(y)$
we can satisfy (\ref{equ1}) up to terms proportional 
to the first power of $y$.
This fixes the half width $y_0$ of the planar N\'eel wall,
\begin{equation} \label{y0}
y_0 = \frac{3\pi}{4 H_0}\,
\sqrt{\frac{K_{\mathrm{Neel}}}{\mu_0 \Delta\chi}},\qquad
K_{\mathrm{Neel}} = \left(
1+\frac{\sr 32}{\sr 9\pi^2}\right) \,K_{33} - K_{11}.
\end{equation}
It is of the order of the magnetic coherence length.
(\ref{y0}) and (\ref{poly1}) are used in (\ref{boundx}),
which now provides the boundary conditions in the $x$ direction.

The approximate solution (\ref{poly1}) could be improved by taking a higher
order polynomial. If it is of the order $y^n$, then (\ref{equ1}) can be
satisfied up to terms proportional to $y^{n-2}$. In the present paper we
shall restrict ourselves to cubic polynomials in $y$, which are sufficient
to reveal our method of obtaining the approximate director field for the
planar soliton.

\section{TILT ANGLE FIELD FOR PLANAR SOLITONS}

Our strategy for solving the non-linear partial differential equation
(\ref{equ}) for the tilt angle $\Phi(x,y)$ proceeds in two steps. First we
apply the polynomial expansion of the tilt angle field in the $y$ coordinate.
After separating the $y$ dependence, we are left with a set of ordinary
differential equations which is solved both numerically and, approximately,
analytically. Of course,
the polynomial expansion in $y$ must satisfy the boundary conditions
(\ref{boundy}). Therefore, up to third order 
(in congruence with the expansion
for the N\'eel wall (\ref{poly1})) it reads
\begin{equation} \label{poly}
\Phi(x,y) = \left( \pm \frac{\Phi_0(x)}{y_0^2} + C(x)\,y\right)\,
(y \mp y_0)^2 \mbox{ mod }\pi, 
\qquad \mbox{for } y \ge 0,\,y \le 0 \mbox{ resp.}
\end{equation}

The polynomial expansion (\ref{poly}) contains two unknown functions
$\Phi_0(x)$ and $C(x)$ that depend on the $x$ coordinate. We can
derive boundary conditions for them by inserting (\ref{poly})
into (\ref{boundx}). This yields (for the positive solitons)
\begin{eqnarray}  \label{boundphi0}
\Phi_0(x=0) =  \frac{\pi}{2}, &\qquad & \Phi_0(x=x_0) = 0, 
\\ \label{boundc}
C(x=0) =  \frac{\pi}{4 y_0^3}, &\qquad & C(x=x_0) = 0.
\end{eqnarray}

Our ansatz (\ref{poly}) is continuous everywhere apart from the
$x$ axis ($y=0$). When crossing the $x$ axis between $x=0$ and
$x= x_0$, a jump in the director orientation from
$+\Phi_0(x)$ to $-\Phi_0(x)$ occurs. 
This is connected to the physical singularity 
of the  disclination line in the center of the defect.
Most significantly, due to the influence of the external magnetic
field the cross-section of the defect core is no more
a point-like object in the $x$-$y$ plane, but it is 
extended to a segment of a straight line of
length $x_0$. However, although the core of the defect is now strip-like (if
we take into account the $z$ direction), one can define its center line. It is
located at $x= x_d$, where $\Phi_0(x=x_d) = \frac{\pi}{4}$,
which gives the largest jump (equal to $\frac{\pi}{2}$) 
in the director orientation at $y=0$. 
At $x \leq 0$ there is no physical singularity, because 
$\Phi_0 = - \frac{\pi}{2}$ is equivalent to $\Phi_0 = + \frac{\pi}{2}$.

The discontinuity of (\ref{poly}) at $y=0$ reflects the fact 
that the continuum approach is no more valid close to the defect
core, where strong gradients of the orientational order are apparent.
On a molecular length scale around the core the mesoscopic 
director looses its physical significance as the average 
molecular orientation. Remarkably, although when using the director
approach we cannot determine the orientational order within the
defect core, our investigation gives hints on a possible elongated
shape of the core of the disclination line in the presence of the magnetic
field. The extension of the defect core ({\em i.e.}~the 
actual value of $x_0$) can
only be determined when including the core energy into the investigation. This
will be performed in the following section.

We now proceed by inserting the third order
polynomial expansion (\ref{poly})
into the equation  (\ref{equ}). By comparison of the
coefficients for the first two powers in the  $y$ coordinate  
({\em i.e.}~$y^0, y^1$) we obtain two ordinary differential equations
for the unknown expansion coefficients $\Phi_0(x)$ and $C(x)$.
It is convenient to  change to a set of dimensionless variables by 
measuring all length scales in units of $y_0$,
\begin{equation} \label{redvar}
x = y_0\,\ov{x}, \qquad \Psi = 2\,\Phi_0, \qquad \Gamma = y_0^3\,C.
\end{equation}
We also introduce the notation
\begin{displaymath}
\frac{\sr 1}{\sr \eta^{2}} = 
1 + \left( 1 + \frac{\sr 9\,\pi^2}{\sr 16} \right) \overline{K}.
\end{displaymath}
The equations for $\Psi$ and $\Gamma$ have the following form
\begin{eqnarray}
& & \frac{\sr 1}{\sr 2}\,(1+ \overline{K}\,\cos\Psi)\,\Psi'' 
+ (1 - \overline{K}\,\cos\Psi)\,(\Psi - 4 \Gamma) 
- \overline{K}\, \left(\frac{\sr 1}{\sr 4} \Psi'\,^2 - 
(\Gamma - \Psi)^2 \right)\,\sin\Psi   \nonumber  \\
& & + 2\, \overline{K}\, ( \Gamma' - \Psi') \,\sin\Psi   +
2\,\overline{K} \,(\Gamma - \Psi)\,\Psi'\,\cos\Psi\,  \nonumber \\
& & - \overline{K}\, (\Gamma -\Psi)\, \Psi' \,\sin\Psi 
- \frac{\sr 1}{\sr \eta^2}\, \sin\Psi = 0,   \label{ord11}
\end{eqnarray}
and
\begin{eqnarray}
& &
\frac{\sr 1}{\sr 2}\,(1+ \overline{K}\, \cos\Psi)\,( \Gamma'' - \Psi'') 
- \frac{\sr 1}{\sr 2}\, \overline{K}\,(\Gamma - \Psi)\, \Psi''\,\sin\Psi  
+ 3\,(1- \overline{K}\cos\Psi)\, \Gamma \nonumber \\
& &
+ \overline{K}\, (\Gamma - \Psi)\, (\Psi - 4 \Gamma)\,\sin\Psi  
- \frac{\sr 1}{\sr 2}\,\overline{K}\, 
\Big( \Psi'\,(\Gamma' - \Psi') - 2\, (\Gamma - \Psi)\,
(\Psi - 4 \Gamma) \Big) \,\sin\Psi  \nonumber \\
& &
+ \overline{K}\, (\Psi' - 4 \Gamma') \, \sin\Psi  
- \overline{K}\, (\Gamma - \Psi) \, 
  \left( \frac{\sr 1}{\sr 4}\, \Psi'^2  - (\Gamma - \Psi)^2 \right)\,
 \cos\Psi
\nonumber \\
& &
+ 2\,\overline{K}\,(\Gamma - \Psi)\, (\Gamma' - \Psi')\, \cos{\Psi}  
- 2\,\overline{K}\,\Psi'\,(\Gamma-\Psi)^2\, \sin\Psi  \nonumber \\
& &
- \overline{K}\, \Psi' \,(\Gamma - \Psi)^2\, \cos\Psi  + 2\,
 \overline{K}\, 
\left( (\Gamma - \Psi)\,(\Gamma' - \Psi') + \frac{\sr 1}{\sr 2}\,\Psi'\,
(\Psi - 4 \Gamma) \right)\, \cos\Psi \nonumber \\
& &
- \overline{K}\, \left( (\Gamma - \Psi)\,(\Gamma' - \Psi') 
+ \frac{\sr 1}{\sr 2}\,
\Psi'\, (\Psi - 4 \Gamma) \right) \, \sin\Psi 
- \frac{\sr  1}{\sr  \eta^2}\, (\Gamma - \Psi)\, \cos\Psi 
= 0.  \label{ord22}
\end{eqnarray}
In (\ref{ord11}) and (\ref{ord22}) $'$ denotes derivatives with respect 
to the dimensionless variable $\ov{x}$.

The set of  ordinary differential equations (\ref{ord11}) and
(\ref{ord22})  becomes  much simpler for the one-constant
approximation ($\overline{K}=0$). 
In this particular case the equations above are
equivalent to the following ones
\begin{eqnarray} \label{ord1}
\Psi'' &=&  8\,\Gamma - 2\,\Psi + 2\,\sin\Psi,
\\ \label{ord2}
\Gamma'' &=& 2\,\Gamma - 2\,\Psi +  2\,
(\Gamma - \Psi)\, \cos\Psi + 2\,\sin\Psi.
\end{eqnarray}
Nevertheless, we shall analyse the set (\ref{ord11}), (\ref{ord22}). It turns
out that a numerical solution and, if $\overline{K}$ is not
too large, also an approximate analytical solution can be obtained.

According to (\ref{boundphi0}) and (\ref{boundc}) 
the boundary conditions (for positive solitons)
now read (with $\overline{x}_0 = x_0/y_0$)
\begin{eqnarray}  \label{boundpsi}
\Psi(\ov{x}=0) = \pi, &\qquad & \Psi(\ov{x}=\ov{x}_0) = 0,
\\ \label{boundgamma}
\Gamma(\ov{x}=0) = \frac{\sr \pi}{\sr 4},
&\qquad & \Gamma(\ov{x}=\ov{x}_0) = 0.
\end{eqnarray}

Eqs.~(\ref{ord11}), (\ref{ord22}), (\ref{boundpsi}), (\ref{boundgamma}) 
define a standard two-point
boundary value problem. It can be solved numerically, for instance
by a {\em shooting method} \cite{Kippenhan,Acton,Numrec}. 
Satisfying the  boundary conditions
at $\ov{x}=0$, the ordinary differential equations
(\ref{ord11}), (\ref{ord22}) are integrated numerically up to $\ov{x}_0$.
The integration constants are adapted iteratively
in order to minimize the discrepancy between the numerical solution 
and the boundary conditions at $\ov{x}_0$. For obtaining
solutions for $\Psi(\ov{x})$ and $\Gamma(\ov{x})$ we used a computer
code from {\em Numerical Recipes} \cite{Numrec}. Our calculations
were performed for parameters corresponding to the liquid
crystalline materials $N$-($p$-methoxybenzylidene)-$p$-
buthylaniline (MBBA) and $p$-azoxyanisole (PAA) ({\em see} section IV).
In these cases the numerical solution almost coincides
with the approximate analytical solution presented below.

An approximate analytical solution of (\ref{ord11}),
(\ref{ord22}) can be achieved, which turns out to be quite accurate, 
as being revealed by a comparison with
the numerical solutions. We start from the observation that the free
energy density of the defect core, which corresponds to a disordered phase, 
is much higher than the typical elastic energy of the nematic. 
Therefore we expect that
the core size $x_0$ is small in comparison with the half-width of the
N\'eel wall $y_0$, {\em i.e.}~$\ov{x}_0 = x_0/y_0 \ll 1$. 
Furthermore, $\Psi$ and
$\Gamma$ change by a finite amount on the interval $[0, \ov{x}_0]$, 
namely by
$\pi$ or $\frac{\pi}{4}$, respectively. 
Therefore, the derivatives $\Psi'$, $\Gamma'$
are of the order $\pi/\ov{x}_0$. They are much larger than $\Psi$ and $\Gamma$.
One cannot a priori exclude that also the second order derivatives are
large, of the order $\pi/\ov{x}_0^{2}$. The approximation consists of keeping
in (\ref{ord11}), (\ref{ord22}) the leading terms only. Then
(\ref{ord11}) reduces to
\begin{equation} \label{reduced1}
(1+\overline{K}\,\cos\Psi)\, \Psi'' 
- \frac{\sr 1}{\sr 2}\, \overline{K}\, \Psi'^2\,\sin\Psi  =0,
\end{equation}
which can immediately be integrated yielding
\begin{equation}
(1+\overline{K}\,\cos\Psi)^{1/2}\, \Psi' = \mbox{ const.}
\end{equation}
In the same approximation  (\ref{ord22}) simplifies to
\begin{eqnarray}
& & (1+\overline{K}\,\cos\Psi)\,(\Gamma'' - \Psi'')
- \overline{K}\,\Psi'\,(\Gamma'-\Psi')\,\sin\Psi \nonumber  \\
& & - \frac{\sr 1}{\sr 2}\, \overline{K}\, 
\,(\Gamma - \Psi)\, \Psi'^2\, \cos\Psi - \overline{K}\,
(\Gamma - \Psi)\, \Psi''\, \sin\Psi = 0. \label{reduced2}
\end{eqnarray}

In addition to the smallness of $\ov{x}_0$ one can also exploit the fact that
the elastic anisotropy $\overline{K}$ can be rather small. For example,
for MBBA its value is 0.11, while for PAA it is  0.42. Moreover, in 
(\ref{reduced1}), (\ref{reduced2})
$\overline{K}$ is multiplied by  the sinus or cosinus
of $\Psi$ -- this effectively diminishes the significance of the terms
proportional to  $\overline{K}$ even further. Therefore, it is natural to look
for solutions of (\ref{reduced1}), (\ref{reduced2}) in the form of
an expansion into powers of $\overline{K}$. Up to first order in
$\overline{K}$ we obtain
\begin{equation} 
\Psi = \pi \left(1- \frac{\ov{x}}{\ov{x}_0}\right) 
- \frac{\sr 1}{\sr 2}\,\overline{K}
\,\sin\left [\pi\,\left(1-\frac{\ov{x}}{\ov{x}_0}\right)\right] 
+ {\cal O}(\overline{K}^2), \label{sol1}
\end{equation}
and
\begin{eqnarray}
\Gamma &=& \frac{\sr \pi}{\sr 4}\, \left(1- \frac{\ov{x}}{\ov{x}_0}\right)
- 2\, \overline{K}\, \sin\left[\pi\,\left(1-\frac{\ov{x}}{\ov{x}_0}
\right)\right] \nonumber \\
& &
+ \frac{\sr 3\pi}{\sr 8}\, \overline{K}\, 
\left(1-\frac{\ov{x}}{\ov{x}_0}\right)\, \left(1 +
\cos\left[\pi\,\left(1-\frac{\ov{x}}{\ov{x}_0}\right)\right]\right) 
+ {\cal O}(\overline{K}^2).  \label{sol2}
\end{eqnarray}
A comparison with the numerical solutions of equations (\ref{reduced1}),
(\ref{reduced2}) shows that the functions in (\ref{sol1}), (\ref{sol2})
yield very good approximations up to  $\overline{K} = 0.6$.

\section{LENGTH OF THE DEFECT CORE}

Up to this stage, the length of the planar soliton $x_0$ (or, equivalently,
$\ov{x}_0 = x_0/y_0$) is unknown. 
We fix it by minimizing the total free energy,
which includes the elastic and magnetic energy of the nematic phase
as well as the energy of the defect core.

Let us first calculate the total nematic energy $F_{\mathrm{nem}}$ 
(per unit length along the $z$ axis) for the soliton extending over the
rectangle $ 0\leq x \leq x_0, \, -y_0 \leq y\leq y_0$, which contains
the core of the defect at $y=0$. 
Outside this rectangle there is the planar N\'eel
wall at $ x \leq0, \, -y_0 \leq y \leq y_0$, and the homogeneous director
orientation 
parallel to the external magnetic field along the three remaining sides of
the rectangle.
Therefore the rectangle contains the total elastic and magnetic energy
of the distorted nematic due to the presence of the defect.
It is given by the integral
\begin{equation}
F_{\mathrm{nem}} = 2\, \int^{x_0}_0 dx\, \int^{y_0}_0 dy \;
{\cal F}_{\mathrm{nem}}[\Phi(x,y)],
\label{int}
\end{equation}
where ${\cal F}_{\mathrm{nem}}$ is given by (\ref{fed}). 
For the tilt angle
$\Phi(x,y)$ we use the approximate solution according to (\ref{poly}),
(\ref{redvar}), (\ref{sol1}) and (\ref{sol2}). The integrals in (\ref{int})
can be calculated by help of a computer algebra system ({\em e.g.}, 
{\sc Maple}).
The result has the following form
\begin{equation}
F_{\mathrm{nem}} = \frac{\sr 1}{\sr 2}\,(K_{11} +K_{33})\,
 \left[ \frac{0.58}{\ov{x}_0} 
- 0.72\, \ov{x}_0 
+ \overline{K}\, \left( \frac{0.19}{\ov{x}_0} + 0.93 
- 11.69\, \ov{x}_0 \right) \right].
\label{toten}
\end{equation}
The terms proportional to $1/\ov{x}_0$ stem from  elastic energy
terms in ${\cal F}_{\mathrm{nem}}$  proportional to $\Phi_{x}^2$. Due to these
terms the elastic energy of a defect with a point-like core would be
infinite, because for such a defect $\ov{x}_0 = 0$.

The expression (\ref{toten}) 
would suggest that $\ov{x}_0$ should be as large as possible
-- then $F_{\mathrm{nem}}$ would be minimal. 
In fact, this is {\em not} the case,
due to the very large free energy stored in the defect core,  where local
transitions into disordered phases may occur. Let us perform an 
estimate of this core energy. 
(Again, it is understood that we consider the energy
per unit length along the $z$ axis.) The energy of the core is due to
large gradients in the orientational order, which appear on a {\em 
molecular}
length scale. Therefore it cannot be expressed in terms of the mesoscopic
director field, but it is related to the molecular interaction potential
across the discontinuity in the tilt angle
on the segment $ 0 \leq x \leq x_0$ of the $x$ axis. This molecular 
interaction energy is small at the beginning and at the end of the core 
where the molecules
on both sides of the segment are almost parallel. However, inside the core
the molecules can even be perpendicular to each other. 
In this latter case the
discontinuity of the tilt angle is equal to $\frac{\pi}{2}$, 
and the separation of centers of mass
of the molecules is of the order $\frac{\sigma_0}{\sqrt{2}}$ 
where $\sigma_0$ denotes
the molecular length. In the present paper we shall be satisfied with a
rough estimate obtained by assuming that the core energy density is
given by (\ref{fed}), when all terms are neglected except for
$\frac{1}{4}\,(K_{11} +K_{33})\,\Phi_{y}^2$, with
$\Phi_y \approx \frac{\pi}{2}/\frac{\sigma_0}{\sqrt{2}}$.  
The width of the core is  taken to
be of the order $\frac{\sigma_0}{\sqrt{2}}$.  This yields an estimate for
the total energy of the core 
\begin{equation}
F_{\mathrm{core}} \approx \frac{\sr 1}{\sr 4}\, 
(K_{11} +K_{33})\, \Phi_y^2  \, x_0
\frac{\sigma_0}{\sqrt{2}} 
= \frac{\pi^2}{8\,\sqrt{2}}\, (K_{11} +K_{33})\,
\frac{y_0}{\sigma}\, \ov{x}_0.
\label{ec}
\end{equation}
The total energy of the planar soliton is then $F = F_{\mathrm{nem}}
+ F_{\mathrm{core}}$. We now insert (\ref{toten}), (\ref{ec}) and 
then minimize $F$ with respect to the reduced core length $\ov{x}_0$.
It is easy to find out that the $\ov{x}_0$ corresponding to
the minimum total energy is given by
\begin{equation}
\ov{x}_0^2 = \frac{0.58 + 0.19\, \overline{K}}{1.74\, y_0/\sigma_0 - 0.72
- 11.69\, \overline{K}}.
\label{l0}
\end{equation}

Let us compute $\ov{x}_0$ for particular nematic materials. For
$N$-($p$-methoxybenzylidene)-$p$-buthylaniline (MBBA)
at  25$^{\circ}$C \cite{Stephen} the elastic constants are
$K_{11} = 6.0\cdot 10^{-12}$ N and $K_{33} = 7.5\cdot 10^{-12}$ N.
The magnetic anisotropy is $\mu_0\,\Delta\chi = 9.7\cdot 10^{-8}$ Vs/Am,
the molecular length $\sigma_0 = 30$ {\AA}.
The magnetic field strength $H_0$ is chosen 500 Oersted, according to a
magnetic flux density $B_0 \equiv \mu_0\,H_0  = 0.05$ T. Then, the elastic
anisotropy is $\overline{K}$ = 0.11. Equation  (\ref{y0}) yields
$y_0$ = 3900 {\AA}. Finally,  $\ov{x}_0^2 \approx 0.0027$,  and $x_0 
\approx 202$ {\AA}.  

For $p$-azoxyanisole (PAA) at 120$^{\circ}$C \cite{Stephen} 
the elastic constants are
$K_{11} = 7.0\cdot 10^{-12}$ N and $K_{33} = 17.0\cdot 10^{-12}$ N.
The magnetic anisotropy is $\mu_0\,\Delta\chi = 12.1\cdot 10^{-8}$ Vs/Am,
the molecular length $\sigma_0 = 20$ {\AA}.
The magnetic field strength $H_0$ is again chosen 500 Oersted.
The elastic anisotropy is $\overline{K}$ = 0.42, $y_0$ = 6850 {\AA},
and finally  $\ov{x}_0^2 \approx 0.0011$,   $x_0 \approx 229$ {\AA}. 

We notice that in both examples $\ov{x}_0^2$ is rather small, indeed. This is
consistent with the assumption leading to the approximate solutions
(\ref{sol1}), (\ref{sol2}). The resulting physical length of the core $x_0$ 
is relatively large and it probably could be seen in appropriate experiments.
The dependence of the nematic, core and total energies on the reduced
length of the defect core is plotted in Figs.~3 (MBBA) and 4 (PAA).

With the determination of the reduced core length $\ov{x}_0$ the calculation
of the director field for the planar soliton is completed.
The tilt angle field is shown in Figs.~5 (positive soliton) and
6 (negative soliton). 
The core line at $y=0$ is clearly visible by
the jump of the tilt angle. However, this picture is somewhat
misleading, because it does not take into account that the
director is an object without arrowhead. For instance, at
$\ov{x}=0,\,\overline{y}=0\pm\,$ 
there is a jump by $\pi$ which in fact means an
orientational change of zero angle, exactly the same as for
$\ov{x}= \ov{x}_0 = 0.052,\,\overline{y}=0\pm$. 
As already stated in the previous section,
due to the periodocity of $\pi$ for tilt angle changes the  
largest orientational jump occurs for $\ov{x}_{d}=0.025,\,
\overline{y}_{d}=0\pm$,
where the tilt angle is $\pm \frac{\pi}{4}$. 
This point can be defined as the center of the core
which is related to the original disclination line
(in three dimensions). 

These particular features become obvious from a
lattice visualization of the director field, which is presented
in Figs.~7 (positive soliton) and 8 (negative soliton). 
(In these figures the $x$ and $y$ dimensions are {\em not}
proportionally scaled.)
Rods of unitary length placed on the sites of a 
rectangular lattice indicate the local orientation. 
The dashed line means the defect core and the small circle
marks the center of the core, according to the previous discussion.

\section{REMARKS}

\begin{enumerate}

\item
The positive and negative planar soliton are distinguished
by the boundary conditions (\ref{boundpsi}) and (\ref{boundgamma}),
but not by the field equations.
Therefore they cover exactly the same area, although their energy content 
is slightly different. The tilt angle field for the negative soliton 
$\Phi^{-}(x,y)$
is obtained from the positive soliton solution $\Phi(x,y)$ 
presented above by a sign inversion of the expansion coefficients:
$\Phi^{-}_{0}(x) = - \Phi_{0}(x)$ and
$C^{-}(x) = - C(x)$.

\item
Expression (\ref{l0}) reveals a rather interesting dependence of 
the reduced core length $\ov{x}_0$ on
the magnetic field $H_0$ which enters through $y_0$ ({\em see} (\ref{y0})).
For weak magnetic fields we have large $y_0$, hence $\ov{x}_0$ is small and
it tends to zero when the magnetic field vanishes. 
However, the physical length
of the core is equal to $x_0 = y_0\,\ov{x}$ 
and it increases as $\frac{1}{\sqrt{H_0}}$
when  the magnetic field decreases. The physical reason is that for a
weaker magnetic field the distance over which the director field can be
reorientated by a given angle is larger. In the case of planar solitons
the required reorientation is such that the tilt angle changes from
$\Phi_{\mathrm{Neel}}(y)$ towards zero.
Of course in this limit the width of
the N\'{e}el domain wall (\ref{y0}) also increases,  as $\frac{1}{H_0}$, 
hence faster.
On the other hand, with increasing  magnetic field $y_0$ decreases  and
$\ov{x}_0$ increases. From inserting (\ref{y0}) into (\ref{l0}) it is
noticed that formally
there is a finite critical value for the magnetic field
at which $\ov{x}_0$ becomes infinite. For MBBA the dependence of 
the reduced and physical core length $\ov{x}_0$ and $x_0$ 
on the reduced magnetic field $h \equiv H[\mbox{Oersted}]/500$ is
\begin{equation}
\ov{x}_0 = 0.55\,\sqrt{\frac{h}{112.8 - h}},\qquad
\frac{x_0[\mbox{\AA}]}{202} = \frac{10.57}{\sqrt{h\,(112.8 - h)}}
\end{equation}
which yields a critical reduced magnetic field $h_{\mathrm{c}}=112.8$, 
corresponding
to a critical flux  density of 5.64 T. (Analogous calculations for PAA
give a somewhat smaller critical flux density of 5.3 T.)
However, it should be remembered that Eq.~(\ref{l0}) has been derived under the
assumption that $\ov{x}_0^2$ is small, so it may become wrong well before 
reaching the critical value of the magnetic field.  

\end{enumerate}

To conclude,  we found an approximate solution
for the director configuration in planar solitons in nematics. 
It is continuous everywhere apart from a strip of finite width.
This points out the possibility of an elongated shape of the defect core 
in disclination lines due to an external magnetic field.

\acknowledgments

The paper is supported in part by KBN grant 2 P03B 095 13.
J.~S.~gratefully acknowledges his individual grant from the
Alexander-von-Humboldt Stiftung.

%
%
\begin{figure}
\caption{Geometry and coordinates for a positive planar soliton
in a nematic liquid crystal.}
\end{figure}
%
%
\begin{figure}
\caption{Geometry and coordinates for a negative planar soliton
in a nematic liquid crystal.}
\end{figure}
%
%
\begin{figure}
\caption{Free energy (per unit length)
$F'$ vs.~core length $\ov{x}_0$ 
for MBBA at 25$^{\circ}$C.
Both quantities are 
in reduced units (dimensionless).
$F' = F/K_{\mathrm{av}}$, $\ov{x}_0 = x_0/y_0$, with units
$K_{\mathrm{av}} \equiv \frac{1}{2}\,(K_{11}+K_{33}) 
= 6.75\cdot 10^{-12}$ N, $y_0 = 0.39\,\mu$m.
Dashed line: analytical solution for the energy
of the nematic phase; rhombs: numerical solution
for the energy of the nematic phase;
dotted line: analytical solution for the energy
of the defect core; crosses: numerical solution
for the energy of the defect core;
solid line: total energy.}
\end{figure}
%
%
\begin{figure}
\caption{Free energy (per unit length)
$F'$ vs.~core length $\ov{x}_0$ 
for PAA at 120$^{\circ}$C.
Both quantities are 
in reduced units (dimensionless).
$F' = F/K_{\mathrm{av}}$, $\ov{x}_0 = x_0/y_0$, with units
$K_{\mathrm{av}} \equiv \frac{1}{2}\,(K_{11}+K_{33}) 
= 12\cdot 10^{-12}$ N, $y_0 = 0.68\,\mu$m.
Dashed line: analytical solution for the energy
of the nematic phase; rhombs: numerical solution
for the energy of the nematic phase;
dotted line: analytical solution for the energy
of the defect core; crosses: numerical solution
for the energy of the defect core;
solid line: total energy.}
\end{figure}
%
%
\begin{figure}
\caption{Tilt angle field for a positive planar soliton
in MBBA at 25$^{\circ}$. Spatial coordinates in reduced
units. $\ov{x} = x/y_0$,  
$\ov{y} = y/y_0$ ($y_0 = 0.39\,\mu$m).} 
\end{figure}
%
%
\begin{figure}
\caption{Tilt angle field for a negative planar soliton
in MBBA at 25$^{\circ}$. Spatial coordinates in reduced
units. $\ov{x} = x/y_0$,  
$\ov{y} = y/y_0$ ($y_0 = 0.39\,\mu$m).} 
\end{figure}
%
%
\begin{figure}
\caption{Lattice visualization of the director configuration 
for a positive planar soliton.} 
\end{figure}
%
%
\begin{figure}
\caption{Lattice visualization of the director configuration 
for a negative planar soliton.} 
\end{figure}

\end{document}